\documentclass[twocolumn,twoside,slac_two]{revtex4}
\usepackage{graphicx}
\usepackage{fancyhdr}
\usepackage{hyperref}
\hypersetup{
    colorlinks=true,
    urlcolor=magenta,
}
\begin{document}

\title{Electron Acceleration Mechanisms in Thunderstorms}

%

\author{S. Celestin}
\affiliation{LPC2E, University of Orleans, CNRS, Orleans, France}
%

\begin{abstract}
Thunderstorms produce strong electric fields over regions on the order of kilometer. The corresponding electric potential differences are on the order of 100 MV. Secondary cosmic rays reaching these regions may be significantly accelerated and even amplified in relativistic runaway avalanche processes. These phenomena lead to enhancements of the high-energy background radiation observed by detectors on the ground and on board aircraft. Moreover, intense submillisecond gamma-ray bursts named terrestrial gamma-ray flashes (TGFs) produced in thunderstorms are detected from low Earth orbit satellites. When passing through the atmosphere, these gamma-rays are recognized to produce secondary relativistic electrons and positrons rapidly trapped in the geomagnetic field and injected into the near-Earth space environment. In the present work, we attempt to give an overview of the current state of research on high-energy phenomena associated with thunderstorms.
\end{abstract}

\maketitle

\thispagestyle{fancy}


\section{INTRODUCTION}

\subsection{High-Energy Radiation Bursts}

\begin{figure}
\includegraphics[width=80mm]{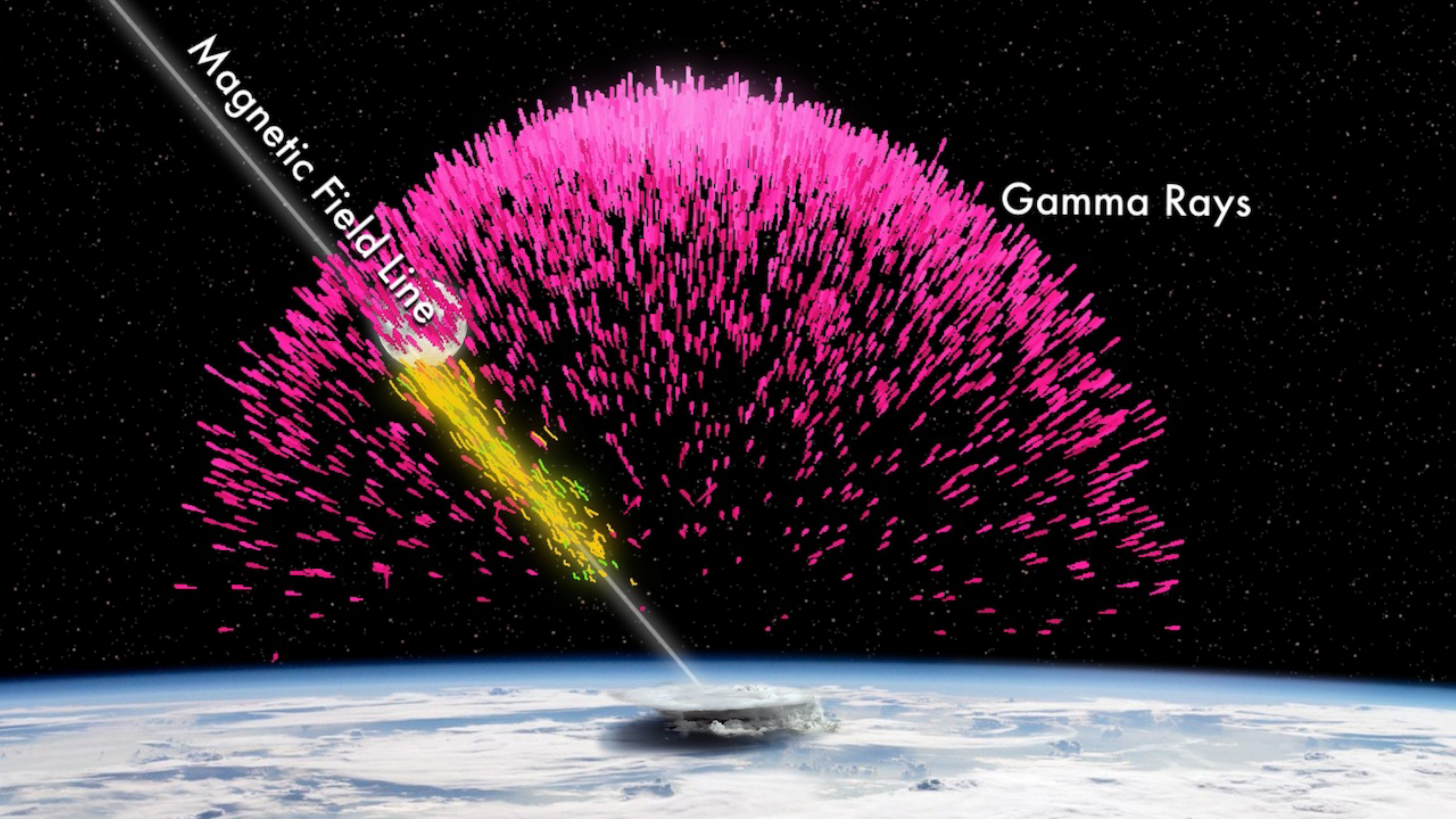}
\caption{Representation of a TGF (pink points) produced by a terrestrial thunderstorm and its associated electron (yellow points) and positron (green points) beams. Credit: NASA/Goddard Space Flight Center/J. Dwyer, Florida Inst. of Technology.}
\label{fig:TGF}
\end{figure}

Terrestrial gamma-ray flashes (TGFs) are bursts of high-energy photons originating from the Earth's atmosphere in association with thunderstorm activity. TGFs were serendipitously discovered in the 1990s by the instrument BATSE on board the Compton Gamma-Ray Observatory (CGRO), which had been originally launched to perform observations of celestial gamma-ray sources \cite{Fishman_Science_1994}. Later on, the detection of these events has been reported using the Reuven Ramaty High Energy Solar Spectroscopic Imager (RHESSI) satellite \cite{Smith_Science_2005}, the Astrorivelatore Gamma a Immagini Leggero (AGILE) satellite \cite{Marisaldi_JGR_2010}, and the Fermi Gamma-ray Space Telescope \cite{Briggs_JGR_2010}.  An illustration of a TGF is given in Figure \ref{fig:TGF}. A few events closely resembling TGFs have also been detected from the ground \cite{Dwyer_GRL_2004c, Tran_JASTP_2015, Dwyer_JGR_2012b, Hare_JGR_2016} and from an aircraft \cite{Smith_JGR_2011}.

Moreover, measurements have correlated TGFs with initial development stages of normal polarity intracloud lightning that transports negative charges upward (+IC) \cite{Stanley_GRL_2006,Shao_JGR_2010, Lu_GRL_2010,Lu_JGR_2011,Cummer_GRL_2015}. Recently, studies focusing on observation and theory of radio emissions believed to be associated with the acceleration processes of electrons producing TGF have provided a new insight in this phenomenon (e.g., see \cite{Cummer_GRL_2011, Connaughton_JGR_2013, Dwyer_JGR_2013, Lyu_GRL_2016b}).

X-ray and gamma-ray bursts have also been recently observed during natural and rocket-triggered lightning discharges usually from the ground (e.g., \cite{Moore_GRL_2001, Dwyer_Science_2003,Dwyer_GRL_2004a,Dwyer_GRL_2005a, Saleh_JGR_2009,Dwyer_JGR_2011c, Mallick_JGR_2012,Schaal_JGR_2012,Schaal_JGR_2014}) but also using airborne detectors \cite{Kochkin_JPD_2015b}. The energy of photons measured in lightning-produced X-ray bursts are typically lower than that in TGFs. 

\subsection{Enhancement of Background Radiation by Thunderstorms}

In addition to radiation bursts, another type of high-energy emission has been observed inside thunderstorms using detectors on board balloons \cite{Eack_JGR_1996, Eack_JGR_2015} and airplanes (e.g., \cite{Parks_GRL_1981, McCarthy_GRL_1985, Kelley_Nature_2015}). These events, so-called gamma-ray glows, correspond to significant enhancements of background radiation that last for more than a few seconds up until a lightning discharge occurs and presumably depletes the charges responsible for the large-scale electric field present in the thunderstorm, which abruptly terminates the gamma-ray glow event. In fact, gamma-ray glows are often observed over a time corresponding to the duration of the presence of the detector in the active region, which is limited by the speed of the aircraft \cite{Kelley_Nature_2015}. Similar emissions have been measured from ground-based detectors (e.g., \cite{Torii_JGR_2002, Torii_GRL_2004, Torii_GRL_2011,Tsuchiya_PRL_2009, Chilingarian_PRD_2010, Chilingarian_AP_2013}) and are usually referred to as Thunderstorm Ground Enhancements.

\subsection{Objectives}

The objectives of the present work are:
\begin{itemize}
\item To present the recent and rapidly growing cross-disciplinary field of research named High-Energy Atmospheric Physics \cite{Dwyer_SSR_2012}.
\item To give insight into the main physical mechanisms and concepts believed to be at play in the acceleration of electrons in thunderstorms.
\item To point out the importance of secondary cosmic rays as high-energy seed particles playing a key role in some of the phenomena presented here.
\item To present future space missions dedicated to the observation of TGFs and the related high-energy particle bursts.
\end{itemize}

\section{Phenomenology of High-Energy Events Associated with Thunderstorms}

\subsection{Terrestrial Gamma-ray Flashes}

The TGF energy spectrum observed by satellites extend from ~10 keV to $>$40 MeV and is typical of bremsstrahlung produced by high-energy electrons accelerated in the atmosphere and colliding with atomic nuclei of air molecules \cite{Ostgaard_JGR_2008}. TGFs duration is shorter than one millisecond \cite{Grefenstette_GRL_2008, Fishman_JGR_2011, Foley_JGR_2014, Marisaldi_JGR_2014, Marisaldi_GRL_2015} (for comparison gamma-ray bursts of cosmic origin (GRBs) have a duration of several seconds or more). When observed from low Earth orbit satellites ($\sim$500 km altitude), they have a typical fluence slightly weaker than $\sim$1 photon/cm$^2$  \cite[e.g.,][]{Briggs_JGR_2010}.

TGFs occur in correlation with lightning discharges inside thunderclouds. Remarkably, they are associated with the most common type of lightning discharges, i.e., intracloud lightning \cite[e.g.,][and references therein]{Cummer_GRL_2015}, occurring in what appears to be common thunderclouds \cite[e.g.,][]{Splitt_JGR_2010,Chronis_BAMS_2016}. The TGF source altitude determined from radio emissions typically ranges between 10 and 14 km \cite{Cummer_GRL_2014}, which is in agreement with the expected spectrum of bremsstrahlung photons produced in this altitude range and transported through the atmosphere up to satellite altitude \cite[e.g.,][]{Dwyer_GRL_2005c, Xu_GRL_2012}. The global TGF occurrence rate is estimated to be 400,000 per year concerning TGFs detectable by Fermi-GBM (Gamma ray Burst Monitor) \cite{Briggs_JGR_2013}, but detailed analysis of satellite measurements \cite{Ostgaard_JGR_2012} and theoretical studies \cite{Celestin_JGR_2015} suggest that it cannot be excluded that TGFs represent the visible part of a ubiquitous process taking place during the propagation of all lightning discharges.

\subsection{Electron and Positron Beams Associated with TGFs}

A remarkable phenomenon associated with TGFs, which was predicted and observed by \emph{Dwyer et al. } \cite{Dwyer_GRL_2008a}, consists in the production of a great number of high-energy electrons and positrons produced by secondary processes accompanying the transport of gamma-rays through the atmosphere. These particles are capable of escaping into space. These events are usually referred to as terrestrial electron beams (TEBs) and represent an unforeseen and not yet quantified source of high-energy electrons and positrons in the EarthÕs radiation belts \cite{Dwyer_GRL_2008a, Briggs_GRL_2011}. They result from the fact that high-energy photons propagating in air are subjected to mainly three collisional processes, which all produce secondary electrons: Photoelectric absorption (main process for energies up to $\sim$30 keV), Compton scattering (main process between $\sim$30 keV and 30 MeV), and electron-positron pair production (main process $>$30 MeV).

\subsection{Gamma-Ray Glows}

Gamma-ray glows associated with thunderstorms are becoming outstandingly interesting, as they are now believed to occur frequently. Indeed, \emph{Kelley et al.} \cite{Kelley_Nature_2015} estimate that a conservative $\sim$8\% of electrified storms produce glows. Although fluxes of photons in glows are much lower than those in TGFs, energy spectra are similar \cite{Kelley_Nature_2015}. Astoundingly, there is now evidence that glows may be of importance to the thundercloud charging mechanisms itself \cite{Kelley_Nature_2015}.

\subsection{Radiation Dose}

\emph{Dwyer et al.} \cite{Dwyer_JGR_2010} have calculated that aircraft passengers finding themselves in an electron beam producing a TGF such as those observed from low-orbit might receive significant radiation doses. Moreover, given the energy of single photons in these events, TGFs are believed to produce neutrons through photonuclear reactions during their transport through the atmosphere \cite{Carlson_JGR_2010a, Kohn_JGR_2015}. \emph{Tavani et al.} \cite{Tavani_NHESS_2013} concluded that TGF-produced neutrons would cause serious hazard on the aircraft electronic equipment. The risk is limited by the fact that pilots try to avoid thunderstorms for other reasons. However, changing routes is not always possible and this is illustrated by the fact that aircraft are regularly hit by lightning discharges. Therefore, a careful risk assessment should be done, especially concerning the cumulated doses received by aircrews.

\subsection{Cosmic Ray Measurements to Probe Thunderstorms}

It is interesting to note that the electric field in thunderstorms is difficult to estimate. Balloon probes for example give local measurements at given times. Recently, the study of cosmic rays has brought a new exciting way to probe thunderstorms. Indeed, \emph{Schellart et al.} \cite{Schellart_PRL_2015} have studied the intensity and polarization patterns of air showers using the radio telescope LOFAR and deduced one integrated component of the large-scale field inside the thunderstorms.

\section{Acceleration of Electrons in Dense Media}

\subsection{Dynamic Friction Force}
\begin{figure}
\includegraphics[width=80mm]{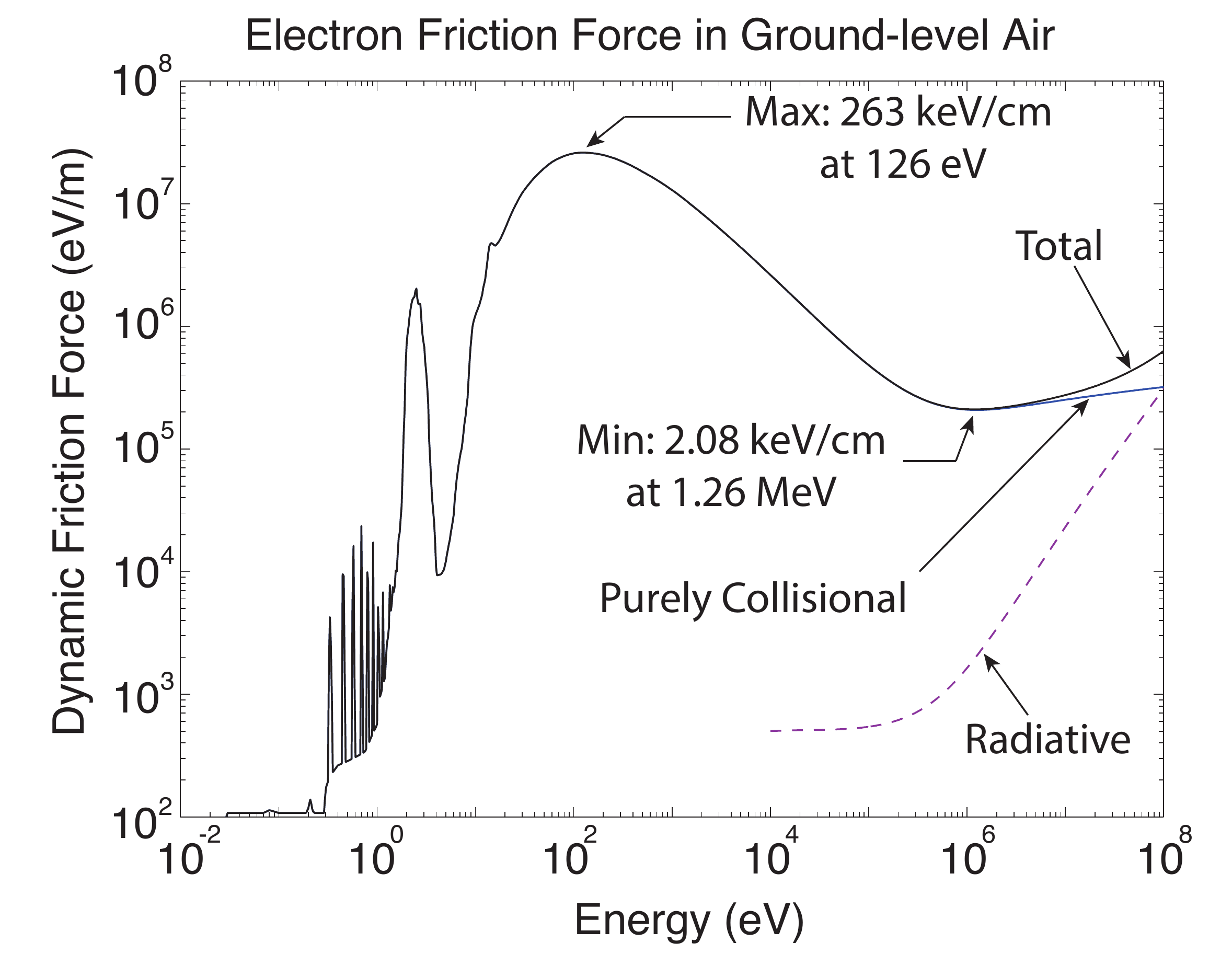}
\caption{Dynamic friction for electrons in air calculated at ground level.}
\label{fig:friction}
\end{figure}

Large scale electric fields (hundreds of meters to kilometers) are usually produced in thunderstorms with amplitudes typically lower than or reaching $2 N/N_0$ kV/cm (e.g., \cite{Marshall_JGR_1995}), where $N$ is the local air density and $N_0$ is the air density at ground level. Free electrons are accelerated under the corresponding electric force, however, because of the high density of the atmosphere at these altitudes (cloud tops altitudes are limited by the tropopause, i.e., $\sim$10--15 km) electrons are usually rapidly slowed down by the collisions with air molecules. From the knowledge of the collisional cross sections and the corresponding energy loss, the dynamic friction (or equivalently the stopping power) of electrons can be calculated. Figure \ref{fig:friction} shows the friction experienced by electrons in units of energy per meter calculated in the atmosphere at ground level (see \cite{ICRU1984,Moss_JGR_2006,Celestin_JPD_2010} for more technical details on the calculation of the friction presented in Figure \ref{fig:friction}).

The complexity of the lower energy part of the friction is caused by the intricacies of inelastic collisional processes between electrons and nitrogen or oxygen molecules while the high-energy part matches very well the Bethe formula with a mean excitation potential estimated for air (e.g., see \cite{ICRU1984}). Since different dedicated models and theories are used in both energy regimes, the seamless representation of electron collisions in air between low- and high-energy ranges is not trivial. It has been the focus of a recent effort, which has proven fruitful in the present field of research \cite{Celestin_JPD_2010, Celestin_JGR_2011, Celestin_JGR_2012, Xu_GRL_2014, Xu_JGR_2015}.

\subsection{Relativistic and Thermal Runaway Electrons}

Because of the amplitude of the dynamic friction in air and its strong increase with the electron energy for energies $<$100 eV, usually, electrons cannot reach high-energy regimes and are quickly thermalized. However, Figure \ref{fig:friction} shows that under an electric field with an amplitude greater than $\sim$2.1 $N/N_0$ kV/cm, high-energy electrons (e.g., secondary cosmic rays) experience an electric force greater than the dynamic friction force. Since they are continuously accelerating and hence separating from the thermal electrons in terms of energy as well as in the configuration space, these electrons are referred to as \emph{runaway electrons}. In fact, mostly due to elastic scattering, the runaway threshold electric field is more precisely $E_{\rm run}\simeq2.8 N/N_0$ kV/cm in air at ground level \citep[e.g.,][]{Dwyer_SSR_2012}. In some occasions, these amplitudes have been observed in thunderstorms \cite{Marshall_JGR_1995}.

While most of ionizing collisions produced by runaway electrons result in low-energy secondary electrons that are not runaway electrons, there is a probability that a secondary runaway electron be produced. This situation occurs if the electric field extends over a region that is larger than the corresponding mean free path of such an ionizing collision. This avalanche mechanism was proposed by \emph{Gurevich et al.} \cite{Gurevich:1992} and is now termed relativistic runaway electron avalanche (RREA)  (see \citep[][and references therein]{Dwyer_SSR_2012}).

However, one can see in Figure \ref{fig:friction} that there is a maximum in the dynamic friction and hence, for extremely strong electric field ($\gtrsim$260 $N/N_0$ kV/cm), thermal electrons can accelerate to high-energy regime \cite{Gurevich:1961}. This process is usually named ``thermal runaway'' \cite{Moss_JGR_2006}. It is worth noting that such high fields are on the order of ten times the conventional breakdown field (field for which the ionization rate equals the electron attachment rate) and because of the strong ionization they would produce, and the corresponding increase of the local conductivity, they cannot be applied locally for durations longer than one nanosecond \cite{Celestin_JGR_2011}. However, filamentary discharges named streamers might be able to sustain these high electric field in a dynamic fashion for longer durations \cite{Moss_JGR_2006,Celestin_JGR_2011}. 

\subsection{Production of Gamma-Rays}

As they are deflected by nuclei of air molecules, electrons produce X- and gamma-rays through bremsstrahlung process. It has been theoretically shown that RREAs in {large-scale} homogeneous electric fields exhibit a robust characteristic exponential high-energy cutoff ($\sim$$7$ MeV) in the electron energy distribution function \cite[e.g.,][]{Dwyer_SSR_2012, Celestin_JGR_2012, Cramer_JGR_2014}. This signature is also present in the associated bremsstrahlung spectra. This property weakly depends on the production altitude and the magnitude of the {large-scale homogeneous} electric field that sustains the RREA \cite{Babich_FizPlaz_2004,Dwyer_GRL_2005c}.

Indeed, average TGF and gamma-ray glow spectra are very well reproduced by RREA-based models  \cite{Dwyer_GRL_2005c, Kelley_Nature_2015}. However, the fluence of TGFs is believed to be too high to be produced by RREAs seeded by natural background radiation or extensive cosmic-ray air showers alone, and as a consequence, relativistic feedback mechanisms have been introduced \citep[e.g.,][]{Dwyer_JGR_2008b,Skeltved_JGR_2014}. If the ambient field is sufficiently strong, exotic large-scale discharges driven by relativistic particles have been predicted to be produced in thunderstorms using numerical models \citep{Liu_JGR_2013}.

\subsection{Theoretical Effort}

X-ray burst produced by cloud-to-ground lightning discharges and observed from the ground are understood to root from the production of thermal runaway electrons \cite{Dwyer_GRL_2005a}, probably caused by the extreme electric fields in streamer discharges during impulsive events occurring in the propagation of negative lightning leaders \cite{Celestin_JGR_2011}. Once thermal runaway electrons are produced, they can accelerate  further in the electric field produced in the vicinity of the lightning leader tip and gain part of the potential drop in the acceleration region \cite{Xu_GRL_2014, Celestin_JGR_2015}.

Gamma-ray glows have similar energy spectra as TGFs \cite{Kelley_Nature_2015}. For this reason, they are believed to be produced by RREAs as well. It seems likely that secondary cosmic rays are accelerated and sometimes even amplified in RREAs in thunderstorms to produce gamma-ray glows. Indeed, fluxes and energy spectra of Thunderstorm Ground Enhancements have been be well-explained by these processes acting on secondary cosmic ray background \cite{Chilingarian_AP_2013}. Moreover, electrons producing gamma-ray glows also produce significant quantities of ions and \emph{Kelley et al.} \cite{Kelley_Nature_2015} have estimated the corresponding ion current. They conclude that, in the glow region, this current is on average comparable with other discharging mechanisms such as lightning and precipitations \cite{Kelley_Nature_2015}. This emphasizes the role of secondary cosmic rays in charging mechanisms of thunderstorms.

Although there seems to be a broad consensus about the physical processes at play in the production of X-ray bursts from lightning discharges and gamma-ray glows, two main theories are usually recognized to be able to explain the occurrence of TGFs in the literature: (1) the direct production of thermal runaway electrons by an ascending lightning leader  at the moment of an impulsive event during its propagation and their further amplification in RREAs developing in the leader-produced field \cite{Moss_JGR_2006, Carlson_JGR_2009, Celestin_JGR_2011, Xu_GRL_2012, Celestin_JGR_2015} and (2) large-scale (kilometers) RREAs in the thundercloud field seeded by either cosmic-ray air showers or thermal runaway electrons produced by a lightning leader \cite[e.g.,][]{Dwyer_JGR_2008b}. Both theories are not mutually exclusive. However, if lightning leaders are to produce TGFs locally (in the field they produce), the two theories lead to different testable predictions \cite[e.g.,][]{Xu_JGR_2015, Celestin_JGR_2015}.

\section{Future Space Missions}

Two missions with TGF measurement as one of their primary goals are planned to be launched within the next few years (2017--2019). Namely, the European Space Agency mission ASIM, which will be placed on board the International Space Station (ISS), and the French space agency (CNES) low-orbit satellite TARANIS. Because they were designed to observe very different sources (e.g., gamma-ray bursts of cosmic origin last several seconds) and given the very high fluxes in TGFs over a timescale of $\sim$100 $\mu$s, spaceborne instruments with TGF-observing capabilities are limited by significant saturation effects \cite[e.g.,][]{Grefenstette_GRL_2008,Tierney_JGR_2013,Marisaldi_JGR_2014}, which change the time dynamics and energy of detected counts as compared to real photons. 

Designed to study TGFs, TARANIS will have very rapid lanthanum-bromide- and plastic-based detectors (XGRE) that should remove, or at least strongly limit, dead time and pile-up effects. The XGRE instrument will also be able to discriminate electrons from photons, which is key to the understanding of terrestrial electron beams (TEBs). XGRE will detect photons from 20 keV to 10 MeV and electrons from 1 MeV to 10 MeV. Other instruments on board TARANIS will detect the associated electromagnetic signals and optical emissions. On board the ISS, ASIM will have imaging capabilities for photons with energies between 10 and 500 keV, and will be able to detect photons with energy up to 40 MeV. ASIM will also be equipped with photometers to detect the related optical emissions.

\section{Conclusions}

\begin{itemize}
\item Since the 1990s, a new field of research named High-Energy Atmospheric Physics has emerged \cite{Dwyer_SSR_2012}.
\item Runaway electrons production, acceleration, and amplification mechanisms are key to understand the physical processes in this new field.
\item Various phenomena have been observed such as TGFs, TEBs, gamma-ray glows, and TGEs.
\item Production mechanisms of some of these phenomena are still up for debate. 
\item The first space missions designed for the study of TGFs and related phenomena will be launched in the coming years: TARANIS (CNES) and ASIM (ESA).
\end{itemize}

\bigskip 
\begin{acknowledgments}
We acknowledge support from the French space agency (CNES) in the framework of the space mission TARANIS.
\end{acknowledgments}

\bigskip 



\end{document}